\RequirePackage{ifpdf}

\documentclass[aps,twocolumn,amsmath,amssymb,preprintnumbers,nofootinbib,tightenlines,nobibnotes]{revtex4-1}

\ifpdf
  \usepackage{color}
\fi

\usepackage{graphicx}
\usepackage{amsmath}
\usepackage{amsfonts}
\usepackage{amssymb}
\usepackage{hyperref}
\usepackage{xcolor}

\newcommand{\bZ}{\mathbb{Z}}
\newcommand{\bC}{\mathbb{C}}

\newcommand{\bR}{\mathbb{R}}

\newcommand{\fatI}{\boldsymbol{I}}

\DeclareMathOperator{\vol}{vol}
\newcommand{\tgd}{\partial_B}
\newcommand{\tgdb}{{{\bar{\partial}_B}}}
\newcommand{\tgdA}{\partial_B^*}
\newcommand{\tgdbA}{\bar{\partial}_B^*}

\DeclareMathOperator{\id}{id}
\DeclareMathOperator{\Hom}{Hom}

\begin{document}

\preprint{RIKEN-MP-74}

\title{Laplace operators on Sasaki-Einstein manifolds}

\author{Johannes Schmude}
\email{johannes.schmude@riken.jp}
\affiliation{Department of Physics, Universidad de Oviedo, 33007, Oviedo, Spain \\
RIKEN Nishina Center, Saitama 351-0198, Japan}

\begin{abstract}
We decompose the de Rham Laplacian on Sasaki-Einstein manifolds
as a sum over mostly positive definite terms. An immediate consequence are
lower bounds on its spectrum. These bounds constitute a supergravity
equivalent of the unitarity bounds in dual superconformal field theories.
The proof uses a generalization of K\"ahler identities to the Sasaki-Einstein case.
\end{abstract}

\maketitle

A textbook result in K\"ahler geometry relates the de Rham with the
Dolbeault Laplacian, $\Delta = 2 \Delta_{\bar{\partial}}$. The main
result of this note is a similar identity for
Sasaki-Einstein manifolds:
\begin{equation}
  \label{eq:main_result}
  \begin{aligned}
    \Delta &= 2 \Delta_\tgdb - \pounds_\xi^2 - 2\imath (n-d^0) \pounds_\xi \\
    &+ 2 L \Lambda + 2 (n-d^0) L_\eta \Lambda_\eta
    + 2 \imath (L_\eta \tgdbA - \tgdb \Lambda_\eta).
  \end{aligned}
\end{equation}
The right hand side of equation \eqref{eq:main_result} features the
Lefschetz operator, the action of the Reeb vector, the tangential
Cauchy-Riemann operator as well as their adjoints. Full definitions
will be given in section \ref{sec:mathematics-proof}. The equation
$\Delta = 2\Delta_{\bar{\partial}}$ can be derived from the K\"ahler
identities, commutators between the Dolbeault and Lefschetz operators
and their adjoints \cite{Huybrechts2005,Voisin2008vol1}. Our proof of
\eqref{eq:main_result} will follow a similar route; we will obtain
K\"ahler-like identities that are valid on Sasaki-Einstein
manifolds. These are summarized in appendix \ref{sec:summary}.

This note is written with two audiences in mind: physicists working in
gauge/string duality or supergravity and mathematicians interested in
Sasaki-Einstein geometry. Therefore we split the discussion into two
separate parts, giving a proof of both \eqref{eq:main_result} and the
identities in section \ref{sec:mathematics-proof}, while discussing
their motivation by and relevance to physics in section
\ref{sec:physics-application}. Readers who want to focus on the
mathematical aspects can ignore section
\ref{sec:physics-application}. Those not interested in the full proof
should read section \ref{sec:mathematics-proof} up to equations
\eqref{eq:adjoints} before skipping ahead to section
\ref{sec:physics-application}.

\section{The proof}
\label{sec:mathematics-proof}

The tangential Cauchy-Riemann operator $\tgdb$ and the associated Kohn-Rossi
cohomology groups $H^{p,q}_\tgdb(S)$ were first introduced in
\cite{Lewy:1956,Kohn1965,Folland:1972}. Given a complex manifold with boundary,
Lewy, Kohn, and Rossi considered under what circumstances 
functions on the boundary can be extended to holomorphic functions in
the bulk. They have to satisfy the projection of the
Cauchy-Riemann equations onto the boundary, hence the name for
$\tgdb$. The Kohn-Rossi cohomology groups feature also in the work by
Yau and collaborators on the complex plateau problem
\cite{Yau1981,Luk2007,Du2012}. This concerns the question when
a real manifold is also the boundary of a complex manifold.

The tangential Cauchy-Riemann operator has properties akin to
those of a Dolbeault operator on a K\"ahler manifold. Yet it is not
the only differential operator on a Sasaki-Einstein manifold with this
characteristic. The simplest example is the basic differential
$\bar{\partial}_{\text{bsc}}$, which arises in the context of basic
forms.\footnote{Our notation is non-standard. The basic differential
  is often denoted as $\tgdb$, with the tangential Cauchy-Riemann
  operator being $\bar{\partial}_b$.}
These are transverse to the Reeb vector and
carry no charge under the Lie derivative along the Reeb. If one drops
the zero-charge condition, it is possible to introduce a transverse
differential $\bar{\partial}_T$. This has been studied by Tievsky
\cite{Tievsky:2008:Analogues}, who obtained transverse K\"ahler
identities, which are a special case of the identities derived in
section \ref{sec:mathematics-proof} and summarized in section
\ref{sec:summary}. Note that we will refer to transverse forms as horizontal.

In what follows, we will give a full proof of
\eqref{eq:main_result} after setting the stage by giving all necessary
definitions. Since the proof is based on the equivalent
considerations in the K\"ahler case, our discussion will follow
\cite{Huybrechts2005,Voisin2008vol1} very closely.

\subsection{Exterior calculus on Sasaki-Einstein manifolds}
\label{sec:exterior-algebra}

Consider a $d = 2n+1$ dimensional Sasaki-Einstein manifold $S$.
Given the Reeb vector $\xi$ and the contact form $\eta$, the
tangent bundle splits as $T S = D \oplus L_\xi$.\footnote{$L_\xi$ is
  the line tangent to $\xi$. In what follows we will set $L_\xi = \xi$
  and $L_\xi^* = \eta$. See section 1 of \cite{Sparks:2010sn} for a
  review of Sasaki-Einstein geometry.}
Furthermore, there is a two-form $J = \frac{1}{2} d\eta$ with $i_\xi J
= 0$. $J$ defines an endomorphism on $TS$ which satisfies $J^2 = -1 +
\xi \otimes \eta$. Since $\eta(D) = 0$, one can decompose the
complexified tangent bundle as $T_\bC S = (\bC \otimes D)^{1,0} \oplus
(\bC \otimes D)^{0,1} \oplus (\bC \otimes \xi)$. This in turn induces
a corresponding decomposition on the complexified cotangent bundle
\begin{equation}\label{eq:decomposition_cotangent_bundle}
  T_\bC^*S = \Omega^{1,0} \oplus \Omega^{0,1} \oplus (\bC \otimes \eta),
\end{equation}
which also extends to the exterior algebra
\begin{equation}\label{eq:decomposition_exterior_algebra}
  \Omega_\bC^* = \bigoplus_{p,q} \Omega^{p,q} \wedge (1\oplus \eta).
\end{equation}
Elements of $\Omega_\bC^*$ that vanish under the action of $i_\xi$ are
called \emph{horizontal}, while those annihilated by $\eta \wedge$ are
\emph{vertical}.

The decomposition \eqref{eq:decomposition_cotangent_bundle} induces a
decomposition of the exterior derivative,
\begin{equation}\label{eq:exterior_d_decomposition}
  d = \tgd + \tgdb + \pounds_\xi \eta \wedge.  
\end{equation}
$\tgdb$ is the \emph{tangential Cauchy-Riemann operator}. 
$\tgd$ and $\tgdb$ satisfy $\{\tgd, \tgdb\} = -2 J \wedge \pounds_\xi$
as well as $\tgd^2 = \tgdb^2 = 0$. The sequence
\begin{equation*}
  \dots \xrightarrow{\tgdb} \Omega^{p,q-1} \xrightarrow{\tgdb}
  \Omega^{p,q} \xrightarrow{\tgdb} \Omega^{p,q+1} \xrightarrow{\tgdb} \dots
\end{equation*}
gives rise to the \emph{Kohn-Rossi cohomology groups}
$H^{p,q}_\tgdb(S)$. Continuing with the
theme of generalizing concepts from K\"ahler geometry to
Sasaki-Einstein manifolds, we define the \emph{Lefschetz operator} $L :
\Omega_\bC^k \to \Omega^{k+2}_\bC$ via $\alpha \mapsto J \wedge \alpha$
and the \emph{Reeb operator} $L_\eta : \Omega_\bC^k \to \Omega_C^{k+1}$
as $\alpha \mapsto \eta \wedge \alpha$.

Introducing the \emph{Hodge star}\footnote{In components
  \begin{equation*}
    \star \alpha_{m_1 \dots m_p} = \frac{\sqrt{g}}{p!} \epsilon_{m_1
      \dots m_{d-p}}^{\phantom{m_1 \dots m_{d-p}} n_1 \dots n_p}
    \alpha_{n_1 \dots n_p}.
  \end{equation*}
}
\begin{equation*}
  \star \bar{\alpha} \wedge \beta = \frac{1}{p!} \bar{\alpha}^{m_1
    \dots m_p} \beta_{m_1 \dots m_p} \vol = \langle \alpha, \beta
  \rangle \vol,
\end{equation*}
allows us to define adjoints for the above operators when acting on $\Omega^k_\bC$:
\begin{equation}
  \label{eq:adjoints}
  \begin{aligned}
    d^* &= (-1)^k \star d \star, \\
    \tgdA &= (-1)^k \star \tgdb \star, \\
    \tgdbA &= (-1)^k \star \tgd \star, \\
    \Lambda &= L^* = \star L \star = J \lrcorner, \\
    \Lambda_\eta &= L_\eta^* = (-1)^{k+1} \star L_\eta \star = i_\xi, \\
    (L_\eta \pounds_\xi)^* &= -\Lambda_\eta \pounds_\xi.
  \end{aligned}
\end{equation}
Recall that on odd-dimensional manifolds $\star$ satisfies $\star\star
= 1$.

When restricted to $D$, the action of $J$ becomes that of an \emph{almost
complex structure} $\mathcal{I}$ which acts as $\mathcal{I}(\alpha) =
J_m^{\phantom{m}n} \alpha_n dx^m$ and $\mathcal{I}(X) = X^m
J_m^{\phantom{m}n} \partial_n$. Of course $\Omega^{1,0} = \{ \alpha
\in \Omega^1 \vert \mathcal{I}(\omega) = \imath \omega \}$.
We also define
\begin{equation*}
  \fatI = \sum_{p,q} \imath^{p-q} \Pi^{p,q},
\end{equation*}
which makes use of the
projection $\Pi^{p,q} : \Omega^*_\bC \to \Omega^{p,q}$.

It will turn out useful to distinguish between the rank of a form on
$\Omega^*_\bC$ and on $\bigwedge^* D^*$. Hence we define
the operator $d^0$ on $\Omega^*_\bC$ via
\begin{equation*}
  d^0 \vert_{\bigwedge^k D^* \wedge \left( 1 \oplus \eta
      \right)} = k \cdot \id.
\end{equation*}
By definition, $d^0$ commutes with $L_\eta$.
A first example of the uses of $d^0$ is given by the notion of
\emph{primitive forms}. $\alpha \in \Omega^*_\bC$ with $d^0 \leq n$ is
primitive if and only if $\Lambda \alpha = 0$. Essentially, the idea of
primitivity on $\bigwedge^* D^*$ is the same as on K\"ahler manifolds,
the contact one-form just comes along for the ride and there is in
no difference between horizontal and vertical forms. We
define $P^k$ as the set of primitive elements of $\bigwedge^k D^*$.

Next we introduce an orthonormal frame $e^i$ on $D^*$. Defining $z^i =
e^{2i-1} + \imath e^{2i}$ and imposing $\mathcal{I} (z^i) = \imath
z^i$, consistency requires that $\mathcal{I} (e^{2i-1}) = -e^{2i}$ and
$\mathcal{I}(e^{2i}) = e^{2i-1}$. Then
\begin{equation*}
  J = \sum_{i}^n e^{2i-1} \wedge e^{2i} = \frac{\imath}{2} \sum_{i}^n
  z^i \wedge \bar{z}^i.
\end{equation*}
Defining $e^{2n+1} = \eta$, one finds $\vol = \vol_{D^*} \wedge
e^{2n+1} = \frac{1}{n!} J^n \wedge \eta$.

In what follows, we will make use of two results concerning the
Hodge star. To begin, assume that $(V, \langle , \rangle)$ is a Euclidean
vector space admiting a decomposition $V = W_1 \oplus W_2$ that is
compatible with the metric $\langle, \rangle$. For simplicity we
assume that $\dim_\bR W_i \in 2\bZ$. The metrics $\langle, \rangle_i$
induce Hodge star operators $\bullet_i$, $i = 1, 2$. Then
$\bigwedge^* V^* = \bigwedge W_1^* \otimes \bigwedge W_2^*$, and for
$\alpha_i \in \bigwedge^{k_i} W_i^*$, the Hodge dual on $\bigwedge^* V^*$, $\bullet$,
threads as  
\begin{equation}
  \bullet (\alpha_1 \otimes \alpha_2) = (-1)^{k_1 k_2} \bullet_1 \alpha_1 \otimes \bullet_2 \alpha_2,
\end{equation}
since ($\beta_i \in W_i$)
\begin{equation*}
  \begin{aligned}
    &\bullet (\alpha_1 \otimes \alpha_2) \wedge (\beta_1 \otimes
    \beta_2) = \langle \alpha_1, \beta_1 \rangle_1 \langle \alpha_2,
    \beta_2 \rangle_2 \vol_1 \vol_2 \\
    &= (-1)^{k_1 k_2} \bullet_1 \alpha_1 \wedge \bullet_2 \alpha_2
    \wedge \beta_1 \wedge \beta_2.
  \end{aligned}
\end{equation*}
One can use identical considerations to decompose the action of
$\star$ on $\Omega^*_\bC$ into seperate operations on $D^*$ and
$\eta$. Introducing a hodge dual $\bullet$ on $D^*$, one finds
\begin{equation}
  \label{eq:Hodge-star-decomposition}
  \star \vert_{\bigwedge^* D^*} = L_\eta \bullet, \qquad
  \star \vert_{\bigwedge^* D^* \wedge \eta} = \bullet (-1)^{d^0} \Lambda_\eta.
\end{equation}

\subsection{Lefschetz decomposition}
\label{sec:lefschetz-decomposition}

The starting point for our discussion of Lefschetz decomposition is
the commutator
\begin{equation}
  \label{eq:Lefschetz-commutator}
  \lbrack L, \Lambda \rbrack = (d^0 - n).
\end{equation}
The proof is via induction in $n$. Consider $d = 3$, $n = 1$. Then $\Omega^*_\bC$ is
spanned by $\{ 1, \eta, \mu_i, J, J \wedge \eta\}$ where $\mu_i \in
\Omega^{1,0} \oplus \Omega^{0,1}$ and both $\mu_i$ are annihilated by
$L$ and $\Lambda$. Then $\Lambda J = 1$ and thus
$\lbrack L, \Lambda \rbrack \vert_{\Omega^0_\bC} = -1$,
$\lbrack L, \Lambda \rbrack \vert_\eta = -1$, $\lbrack L, \Lambda \rbrack \vert_{D^* = 0}$,
$\lbrack L, \Lambda \rbrack \vert_{\Omega^{1,1}} = 1$, and 
$\lbrack L, \Lambda \rbrack \vert_{\Omega^3_\bC} = 1$. Hence
\begin{equation*}
  \lbrack L, \Lambda \rbrack \vert_{\bigwedge^k D^* \wedge (1 \oplus
    \eta)} = (k - 1), \qquad k = 0, 1, 2,
\end{equation*}
as claimed. The induction then proceeds as in \cite{Huybrechts2005}

\eqref{eq:Lefschetz-commutator} generalizes to
\begin{equation}
  \label{eq:Lefschetz-commutator_i}
  \lbrack L^i, \Lambda \rbrack \vert_{\bigwedge^k D^* \wedge (1\oplus
    \eta)} = \imath (k-n+i-1) L^{i-1}.
\end{equation}
Again the proof is a copy of that in \cite{Huybrechts2005}.

To proceed we follow \cite{Voisin2008vol1}. Restricting to
$\bigwedge^* D^*$ one can copy all results from proposition $6.20$ to
lemma $6.24$. The most important of these results is \emph{Lefschetz
  decomposition}. Given $\alpha \in \bigwedge^k D^*$, there is a
unique decomposition
\begin{equation*}
  \alpha = \sum_r L^r \alpha_r, \qquad \alpha \in P^{k-2r}.
\end{equation*}
The decomposition is compatible with the
bidigree decomposition and with the decomposition into horizontal and
vertical components. Moreover,
\begin{equation}
  \label{eq:Lefschetz-isomorphism}
  L^{n-k} : \bigwedge^k D^* \to \bigwedge^{2n-k} D^*
\end{equation}
is an isomorphism and the primitivity condition is equivalent to
$L^{n-k+1} \alpha = 0$.

The Lefschetz decomposition becomes incredibly useful when used
together with the Bidigree decomposition, equation
\eqref{eq:Hodge-star-decomposition} and the identity
\begin{equation}
  \label{eq:mysterious_identity}
  \begin{aligned}
    \forall \alpha \in P^k, \quad
    \bullet L^j \alpha &= F(n, j, k) L^{n-k-j} \fatI (\alpha), \\
    F(n,j,k) &= (-1)^{\frac{k(k-1)}{2}} \frac{j!}{(n-k-j)!}.
  \end{aligned}
\end{equation}
Since no differential operators are involved and $\alpha \in
\bigwedge^k D^*$, one can copy the proof in \cite{Huybrechts2005}
after adjusting for conventions. Once the dust settles, the only
difference is in the $k$-dependent prefactor.

\subsection{K\"ahler-like identities}
\label{sec:kahler-like-identities}

We are finally in a position to make use of the previous results and
calculate the (anti-) commutators. The results are in summarized in table
\ref{tab:identities}. A number of identities are fairly obvious:
\begin{equation*}
  \begin{aligned}
    0 &= \lbrack \tgd, L \rbrack = 
    \lbrack \tgdb, L \rbrack = 
    \lbrack \tgdA, \Lambda \rbrack =
    \lbrack \tgdbA, \Lambda \rbrack \\
    &= \lbrack L, L_\eta \rbrack = 
    \lbrack \Lambda, \Lambda_\eta \rbrack =
    \lbrack L_\eta, \Lambda \rbrack.  
  \end{aligned}
\end{equation*}
One finds $\{ L_\eta, \Lambda_\eta \} = 1$ by direct calculation using
the decomposition $\alpha = \alpha_H + L_\eta \alpha_V$. Finally,
$\lbrack d^0, \tgd \rbrack = \tgd + L \Lambda_\eta$.

The most involved calculation is that of the commutator
\begin{equation}\label{eq:most_difficult_commutator}
  \lbrack \Lambda, \tgdb \rbrack = -\imath\tgdA + \imath L_\eta
  \Lambda + (n-d^0) \Lambda_\eta.
\end{equation}
Before we turn to the proof, let us try to interpret this result as a
generalization of the K\"ahler case
$\lbrack \Lambda, \bar{\partial} \rbrack = - \imath \partial^*$. The naive guess 
$\lbrack \Lambda, \tgdb \rbrack \overset{?}{=} - \imath \tgdA$ cannot be correct
since the left hand side maps $\lbrack \Lambda, \tgdb \rbrack:
\bigwedge^* D^* \to \bigwedge^* D^*$ while $\tgdA : \bigwedge^*
D^* \to \bigwedge^* D^* \wedge (1 \oplus \eta)$. Similarly, the right
hand side annihilates $\eta$ while the left hand side does not. One
can guess the correct result by considering the action of both sides
on $J$ and $\eta$, adding suitable terms on the right hand side to
achieve equality.

The proof of \eqref{eq:most_difficult_commutator} is once again an
elaboration on the proof for K\"ahler manifolds in
\cite{Huybrechts2005}. Let us first consider horizontal forms. Here,
it is sufficient to explicitly evaluate the action of
\eqref{eq:most_difficult_commutator} $L^i \alpha$ for $\alpha \in
P^k$; the result will generalize for generic elements of $\bigwedge^*
D^*$ due to Lefschetz decomposition. Furthermore one applies Lefschetz
decomposition to $\tgdb \alpha = \alpha_0 + L \alpha_1 + L^2 \alpha_2
+ \dots$. We have $\alpha \in P^k$ and thus $0 = \sum_j L^{n-k+1+j}
\alpha_j$ and finally $L^{n-k+1+j} \alpha_j = 0$. Using equation
\eqref{eq:Lefschetz-isomorphism} it follows that most of the
$\alpha_j$ vanish and $\tgdb\alpha = \alpha_0 + L \alpha_1$.

Using \eqref{eq:Lefschetz-commutator_i} one finds
\begin{equation*}
  \lbrack \Lambda, \tgdb \rbrack L^i \alpha = -i L^{i-1} \alpha_0 -
  (k+i-n-1) L^i \alpha_1.    
\end{equation*}
Similarly, using $\tgdb \fatI(\alpha) = \imath \fatI
(\tgdb\alpha)$ and $\fatI^2 (\bigwedge^k D^*) = (-1)^k$ as well as
\eqref{eq:Hodge-star-decomposition} and \eqref{eq:mysterious_identity}
\begin{equation*}
  \begin{aligned}
    \star \tgdb \star L^i \alpha &= \imath (-1)^{k^2} \lbrack \Lambda, \tgdb \rbrack L^i \alpha -
    (-1)^k L_\eta \lbrack L^i, \Lambda \rbrack \alpha.
  \end{aligned}
\end{equation*}
Finally,
\begin{equation*}
  \lbrack \Lambda, \tgdb \rbrack \vert_{\bigwedge^* D^*} = -\imath
  \tgdA + \imath L_\eta \Lambda.    
\end{equation*}

To study vertical forms, we consider $L_\eta L^i \alpha$. Again
$\alpha \in P^k$ and $\tgdb \alpha = \alpha_0 + L \alpha_1$. Then
\begin{equation*}
  \begin{aligned}
    \lbrack \Lambda, \tgdb \rbrack L_\eta L^i \alpha &= i L_\eta
    L^{i-1} \alpha_0 + (k+i-n-1) L_\eta L^i \alpha_1 \\
    &+ \lbrack n-(2i+k)\rbrack L^i \alpha.
  \end{aligned}
\end{equation*}
Note that $2i+k$ is the degree of $L^i \alpha$. Furthermore,
\begin{equation*}
  \begin{aligned}
    \star \tgdb \star L_\eta L^i \alpha &= (-1)^{k^2+1} \imath \\
    &\times \lbrack
    i L_\eta L^{i-1} \alpha_0 + (k+i-n-1) L_\eta L^i \alpha_1 \rbrack.
  \end{aligned}
\end{equation*}
In total,
\begin{equation*}
  \lbrack \Lambda, \tgdb \rbrack (L_\eta L^i \alpha) = \{ -\imath \tgdA
  + \lbrack n - (2i+k) \rbrack \Lambda_\eta \} (L_\eta L^i \alpha).
\end{equation*}
Since $L_\eta \Lambda (L_\eta L^i \alpha) = 0$, we can add or
subtract $\imath L_\eta \Lambda$. Therefore it is consistent to
combine the results on horizontal and vertical forms into the
overall result \eqref{eq:most_difficult_commutator}. An identical
calculation or complex conjugation give $\lbrack \Lambda, \tgd
\rbrack$. This completes the proof.

We can compute the computator of the adjoints ($\alpha \in \Omega^p_\bC$):
\begin{equation*}
  \lbrack L, \tgdA \rbrack \alpha = (-1)^p \lbrack -\imath
  \star \tgdA \star + \imath \star L_\eta \Lambda \star +
  \star (n-d^0) \Lambda_\eta \star \rbrack \alpha_p.
\end{equation*}
With $\star (n-d^0) \star = (d^0 - n)$, $\star \tgdA \star
\alpha = (-1)^{p+1} \tgdb\alpha$, and
$\star L_\eta \Lambda \star \alpha = (-1)^{p+1} \Lambda_\eta L
\alpha$ one finds
\begin{equation*}
  \begin{aligned}
    \lbrack L, \tgdA \rbrack &= \imath \tgdb - \imath
    \Lambda_\eta L + (d^0 - n) L_\eta, \\
    \lbrack L, \tgdbA \rbrack &= -\imath \tgd + \imath
    \Lambda_\eta L + (d^0 - n) L_\eta.
  \end{aligned}
\end{equation*}

The calculation of the anticommutator $\{ \Lambda_\eta, \tgdb \}$ is
considerably simpler. Consider again $\alpha \in P^k$ with $\tgdb
\alpha = \alpha_0 + L \alpha_1$. Then $\Lambda_\eta \tgdb \alpha = 0$
and $\tgdb \Lambda_\eta \alpha = 0$. The next step is only slightly
more complicated: $\Lambda_\eta \tgdb L_\eta \alpha = -\tgdb \alpha$,
$\tgdb \Lambda_\eta L_\eta \alpha = \tgdb \alpha$ and thus
$\{ \Lambda_\eta, \tgdb \} = 0$. Similarly $\{ \Lambda_\eta, \tgd \} =
0$. The adjoint anticommutator vanishes, too.

This concludes the calculation of the identities. The (anti-) commutators
allow us to express $\Delta = d^* d + d d^*$ in terms of
$\Delta_\tgdb = \tgdbA \tgdb + \tgdb \tgdbA$.
The decomposition \eqref{eq:exterior_d_decomposition} yields
\begin{equation*}
  \Delta = \Delta_{\tgd} + \Delta_\tgdb + \{\tgd, \tgdbA\} +
    \{ \tgdb, \tgdA \} - \pounds_\xi^2.
\end{equation*}
Then, using 
$\{\tgd, \tgdbA\} = \{\tgd, L_\eta \Lambda\} + \imath \tgd \Lambda_\eta$
one shows that
\begin{equation*}
  \Delta_{\tgd} = \Delta_\tgdb - 2\imath (n-d^0)
  \pounds_\xi + \{\tgd-\tgdb, L_\eta \Lambda\} - \imath (\tgd + \tgdb)
  \Lambda_\eta,
\end{equation*}
which leads to
\begin{equation*}
  \Delta = 2 \Delta_\tgdb - 2\imath (n-d^0)
  \pounds_\xi - \pounds_\xi^2 + 2 \{ \tgd, L_\eta \Lambda \} - 2
  \imath \tgdb \Lambda_\eta.  
\end{equation*}
Application of
$\{ \tgd, L_\eta \Lambda \} = \imath L_\eta \tgdbA + (n-d^0)
L_\eta \Lambda_\eta + L\Lambda$
completes the proof of \eqref{eq:main_result}.

\subsection{Beyond K\"ahler identities?}
\label{sec:discussion-maths}

Since we found Sasaki-Einstein equivalents of both
$\Delta = 2\Delta_{\bar{\partial}}$ and the K\"ahler identities, it is
tempting to ask how much more of K\"ahler geometry can be
generalized. For example, since $\Delta_{\bar{\partial}}$ is
self-adjoint and elliptic, one can show that $\Omega^k_\bC =
\mathcal{H}^k \oplus \Delta_{\bar{\partial}} (\Omega^k_\bC)$ which
implies Hodge's theorem. Similarly, the relation between the de Rham
and Hodge Laplacians allows for an isomorphism between the respective
spaces of harmonic forms. However, it turns out that $\Delta_{\tgdb}$
is not elliptic. We will sketch the calculation leading to this
result. Recall that $\Delta_{\tgdb}$ is elliptic if the symbol
$\sigma_{\Delta_{\tgdb}} \in \Hom(\Omega^k_\bC, \Omega^k_\bC) \otimes
S^2(T^*S)$ maps any non-zero $\omega \in T^*S$ to an automorphism on
$\Omega^k_\bC$. When calculating the symbol one essentially keeps only
those terms of $\Delta_{\tgdb}$ that are of highest order in
derivatives. In the context of the tangential Cauchy-Riemann operator,
this means that $\tgd$ and $\tgdb$ can be taken to be anticommuting
and that the overall result is essentially the same as for the symbol
of the Dolbeault Laplacian on a K\"ahler manifold, provided one
substitutes $\partial_{z^i} \mapsto \partial_{z^i} -
\eta(\partial_{z^i}) \pounds_\xi$. Therefore,
$\sigma_{\Delta_{\tgdb}}(\xi) = 0$ and $\Delta_{\tgdb}$ is not
elliptic. Tievsky's discussion of a transverse Laplacian $\Delta_T$
arrives at a similar result. In that case, it turns out that $\Delta_T
- (\Lambda_\eta d)^2$ is elliptic. A similar result should hold here,
possibly after replacing $\Lambda_\eta d$ with $\pounds_\xi$
\cite{Tievsky:2008:Analogues}. El Kacimi-Alaoui has studied elliptic
operators acting on basic forms \cite{ElKacimi1990}.

\section{Motivation and applications}
\label{sec:physics-application}

Both equation \eqref{eq:main_result} as well as the identities in
appendix \ref{sec:summary} find application in the AdS/CFT
correspondence. Freund-Rubin compactification on Sasaki-Einstein
manifolds yields supergravity duals of superconformal field
theories
\cite{Kehagias:1998gn,Klebanov:1998hh,Acharya:1998db,Morrison:1998cs,Martelli:2004wu}.
The AdS/CFT
dictionary links the conformal energy of SCFT operators to the spectrum of
$\Delta$, their $R$-charge to that of the Lie-derivative along the
Reeb vector, $\pounds_\xi$.
The conformal energy, $R$-charge, and spin of any SCFT operator
have to satisfy the unitarity bounds \cite{Minwalla:1997ka}, which should be reflected on
the supergravity side in the spectrum of $\Delta$. We will argue shortly
that equation \eqref{eq:main_result} allows us to re-derive the
unitarity bounds from supergravity when considered in conjunction with
the calculations in \cite{Eager:2012hx,Eager:2013mua}. 

First, note that the K\"ahler-like identities allow for a study of
the eigenmodes of $\Delta$. In the case where the Sasaki-Einstein
manifold has a coset structure, this has been done using harmonic analysis
\cite{Ceresole:1999ht}. \cite{Eager:2012hx,Eager:2013mua} obtained the
structure of the Kaluza-Klein spectrum of generic Sasaki-Einstein
manifolds using a construction similar to that in
\cite{Pope:1982ad}, which can be nicely summarized in terms of the
identities in appendix \ref{sec:summary}:
Given any eigen $k$-form $\omega$ of $\Delta$, one diagonalizes the
action  of $\Delta$ on the $k+1$-forms
$\{ \tgd \omega, \tgdb \omega, L_\eta \omega, L \omega, \tgd\tgdb
\omega, \dots \}$. The resulting eigenstates fill out representations
of the superconformal algebra, equivalence classes in Kohn-Rossi cohomology groups
$H^{p,q}_{\tgdb}(SE)$ correspond to short multiplets. Whereas the
original calculations were based on a rather tiresome direct approach,
the methods developed in this note are expected to simplify that kind
of anlysis considerably.

With this in mind,
we turn to the spectral problem for $\Delta$. Consider a
$k$-form $\omega$ with $\pounds_\xi \omega = \imath q$, $q \geq 0$, 
and $d^0 \leq n$. All terms on the right hand side of
\eqref{eq:main_result} are positive definite except for the mixed term 
$M = \imath (L_\eta \tgdbA - \tgdb \Lambda_\eta) = N + N^*$. $M$ is
self-adjoint and its spectrum is real. Moreover, $N^2 = 0$ and
$N(\bigwedge^* D^*) \subset \bigwedge^* D^* \wedge \eta$ and
$N(\bigwedge^* D^* \wedge \eta) = 0$. That is, $N$ maps horizontal to
vertical forms and annihilates the latter. $N^*$ behaves accordingly
and it follows that $\langle \omega, M \omega \rangle$ vanishes if
$\omega$ is horizontal or vertical. This is also the case if $\omega$
is neither horizontal nor vertical yet holomorphic.\footnote{In the
  remainder of this discussion, the term \emph{holomorphic} is meant
  in respect to the tangential Cauchy-Riemann operator $\tgdb$.}
As long as we restrict to one of these cases, \eqref{eq:main_result}
takes the form of a bound on the spectrum of $\Delta$.

This was conjectured and partially shown in the context of the calculations of the
superconformal index \cite{Kinney:2005ej,Bhattacharya:2008zy} in
\cite{Eager:2012hx,Eager:2013mua}. Here, the
spectrum was constructed from primitive elements of $\Omega^{p,q}$. For such
forms, \eqref{eq:main_result} implies
\begin{equation}\label{eq:the_old_bound}
\Delta \geq q^2 + 2q (n-d^0)
\end{equation}
with equality if and only if $\tgdb 
\omega = \tgdbA \omega = 0$. In the K\"ahler case, the latter of these
is implied by transversality --- $d^* \omega = 0$. Here however,
$d^* \omega = 0$ leads only to the vanishing of the horizontal
component of $\tgdbA \omega$. Indeed,
\begin{equation*}
  \tgdA \omega = \imath L_\eta \Lambda \omega, \quad
  \tgdbA \omega = - \imath L_\eta \Lambda \omega,
\end{equation*}
which vanishes since $\omega$ was assumed to be primitive. Assuming
that every element of $H^{p,q}_\tgdb(S)$ has a representative closed
under $\tgdbA$, the bound \eqref{eq:the_old_bound} is saturated on the
elements of $H^{p,q}_\tgdb(S)$. These are the forms that correspond to
the short multiplets in the SCFT, and \eqref{eq:the_old_bound}
together with the expressions for the derived eigenmodes of $\Delta$
given in \cite{Eager:2012hx,Eager:2013mua} allows to recover the
unitarity bounds from supergravity. Note that \eqref{eq:the_old_bound}
and a precursor to \eqref{eq:main_result} were already conjectured in
those references. Furthermore, the appendix of \cite{Eager:2013mua}
contains an argument that every element of $H^{p,q}_\tgdb(S)$ is
either primitive, carrying zero charge, or both. For the cases of
interest in the context of that paper it turned out that all elements
are primitive.

A further application of \eqref{eq:main_result} is the stability
analysis of Pilch-Warner solutions in
\cite{Pilch:2013gda}. In the absence of general theorems concerning
Laplace operators on Sasaki-Einstein manifolds, the authors
constructed explicitly examples of primitive, basic $(1,1)$-forms whose existence
renders these solutions perturbatively unstable.
Assuming that the calculations in \cite{Pilch:2013gda} generalize to
generic transverse forms, our results might allow for a continuation of their
analysis to manifolds where explicit constructions are not feasible.

Finally, it would be interesting to extend the results
presented here beyond the Sasaki-Einstein case. As long as there is a
dual SCFT, there is a unitarity bound meaning that there should be
some equivalent of \eqref{eq:main_result} or at least
\eqref{eq:the_old_bound}.

\begin{table*}[bhtp]
  \centering
  \begin{tabular}{|c|ccccccc|}
    \hline
    & $L_\eta$ & $\Lambda_\eta$ & $L$ & $\Lambda$ 
    & $\tgdA$ & $\tgdbA$ & $\tgdb$ \\
    \hline
    $\tgd$ & $\{ \tgd, L_\eta \} = L$ & $\{ \tgd, \Lambda_\eta \} = 0$
    & $\lbrack \tgd, L \rbrack = 0$ 
    & $\lbrack \tgd, \Lambda \rbrack$ 
    & $\Delta_{\tgd}$ 
    & $\{ \tgd, \tgdbA \}$ 
    & $\{\tgd, \tgdb \} = -2 L \pounds_\xi$ \\
    $\tgdb$ & $\{\tgdb, L_\eta \} = L$ & $\{\tgdb, \Lambda_\eta \} = 0$
    & $\lbrack \tgdb, L \rbrack = 0$ 
    & $\lbrack\tgdb, \Lambda\rbrack$
    & $\{ \tgdb, \tgdA \}$
    & $\Delta_\tgdb$ & \\
    $\tgdbA$ & $\{\tgdbA, L_\eta\} = 0$
    & $\{\tgdbA, \Lambda_\eta \} = \Lambda$
    & $\lbrack\tgdbA, L\rbrack$
    & $\lbrack \tgdbA, \Lambda \rbrack = 0$
    & $\{\tgdbA, \tgdA\} = 2 \Lambda \pounds_\xi$
    & & \\
    $\tgdA$ & $\{\tgdA, L_\eta \} = 0$ 
    & $\{\tgdA, \Lambda_\eta \} = \Lambda$
    & $\lbrack \tgdA, L \rbrack$
    & $\lbrack \tgdA, \Lambda \rbrack = 0$ & & & \\
    $\Lambda$ & $\lbrack\Lambda, L_\eta\rbrack =0$
    & $\lbrack \Lambda, \Lambda_\eta \rbrack = 0$
    & $\lbrack\Lambda, L\rbrack = (n-d^0)$ & & & & \\
    $L$ & $\lbrack L, L_\eta\rbrack = 0$ & $\lbrack L, \Lambda_\eta \rbrack = 0$
    & & & & & \\
    $\Lambda_\eta$ & $\{\Lambda_\eta, L_\eta \} = 1$ & & & & & & \\
    \hline
  \end{tabular}
  \caption{The K\"ahler-like identities}
  \label{tab:identities}
\end{table*}

\section*{Acknowledgements}
\label{sec:acknowledgements}

I would like to thank Richard Eager and Yuji Tachikawa for the
collaborations and discussions that lead to the results presented
here. I would also like to thank Jyotirmoy Bhattacharya for
discussions and James Sparks for correspondence related to this
project. Finally, I would like to thank the organizers of Strings 2013
and the joint YITP/KIAS workshop where most of this paper was written.

\appendix

\section{The identities}
\label{sec:summary}

Table \ref{tab:identities} summarizes the various (anti-) commutators. The more
involved ones that do not fit in the table are listed in equation
\eqref{eq:long_se_identities}.

\begin{equation}\label{eq:long_se_identities}
  \begin{aligned}
    \lbrack \tgd, \Lambda \rbrack &= -\imath \tgdbA + \imath
    L_\eta \Lambda - (n - d^0) \Lambda_\eta \\
    \lbrack\tgdb, \Lambda\rbrack &= \imath \tgdA - \imath
    L_\eta \Lambda - (n-d^0) \Lambda_\eta, \\
    \lbrack \tgdA, L \rbrack &= -\imath\tgdb + \imath
    \Lambda_\eta L - (d^0 - n) L_\eta, \\
    \lbrack \tgdbA, L \rbrack &= \imath\tgd - \imath
    \Lambda_\eta L - (d^0 - n) L_\eta, \\
    \{ \tgd, \tgdbA \} &= \imath(L_\eta \tgdbA + \tgd
    \Lambda_\eta) + (n-d^0) L_\eta \Lambda_\eta + L \Lambda, \\
    \{ \tgdb, \tgdA \} &= -\imath (L_\eta \tgdA + \tgdb
    \Lambda_\eta) + (n-d^0) L_\eta \Lambda_\eta + L\Lambda.
  \end{aligned}
\end{equation}

\bibliography{ref}

\end{document}